\documentclass[slac_one]{revtex4}
\usepackage{graphicx}
\usepackage{epsfig}
\usepackage{fancyhdr}
\usepackage{amsfonts}
\usepackage{amsmath}
\usepackage{amssymb}
\usepackage[dvips]{color}
\usepackage{colordvi,psfrag}
\makeatletter
\def\citer{\@ifnextchar [{\@tempswatrue\@citexr}{\@tempswafalse\@citexr[]}}
 
\def\@citexr[#1]#2{\if@filesw\immediate\write\@auxout{\string\citation{#2}}\fi
  \def\@citea{}\@cite{\@for\@citeb:=#2\do
    {\@citea\def\@citea{--\penalty\@m}\@ifundefined
       {b@\@citeb}{{\bf ?}\@warning
       {Citation `\@citeb' on page \thepage \space undefined}}%
\hbox{\csname b@\@citeb\endcsname}}}{#1}}
\makeatother

\def\citere#1{\mbox{Ref.~\cite{#1}}}

\def\xtilde#1{%
  \setbox0\hbox{$\tilde#1$}%
  \rlap{\raise\ht0\hbox{\tiny$_{\,(\;\,)}$}}%
  \tilde#1%
}

\newcommand{\MstL}{M_{\tilde{t}_L}}
\newcommand{\MstR}{M_{\tilde{t}_R}}

\newcommand{\At}{A_t}
\newcommand{\Ab}{A_b}

\newcommand{\Xt}{X_t}

\newcommand{\Msusy}{M_{\rm SUSY}}

\newcommand{\mh}{m_h}

\newcommand{\gl}{\tilde{g}}

\newcommand{\mgl}{m_{\tilde{g}}}

\newcommand{\tsf}{\theta\kern-.20em_{\tilde{f}}}
\newcommand{\tsfp}{\theta\kern-.20em_{\tilde{f}\prime}}
\newcommand{\tsq}{\theta\kern-.15em_{\tilde{q}}}

\newcommand{\VL}{\left( \begin{array}{c}}
\newcommand{\VR}{\end{array} \right)}
\newcommand{\ML}{\left( \begin{array}{cc}}
\newcommand{\MLd}{\left( \begin{array}{ccc}}
\newcommand{\MLv}{\left( \begin{array}{cccc}}
\newcommand{\MR}{\end{array} \right)}

\newcommand{\tb}{\tan \beta}

\newcommand{\tev}{\,\, {\rm TeV}}
\newcommand{\gev}{\,\, {\rm GeV}}

\newcommand{\BC}{\begin{center}}
\newcommand{\EC}{\end{center}}
\newcommand{\BE}{\begin{equation}}
\newcommand{\EE}{\end{equation}}
\newcommand{\BEA}{\begin{eqnarray}}
\newcommand{\BEAnn}{\begin{eqnarray*}}
\newcommand{\EEA}{\end{eqnarray}}
\newcommand{\EEAnn}{\end{eqnarray*}}
\newcommand{\non}{\nonumber}
\newcommand{\id}{{\rm 1\kern-.12em
\rule{0.3pt}{1.5ex}\raisebox{0.0ex}{\rule{0.1em}{0.3pt}}}}
\newcommand{\lsim}
{\;\raisebox{-.3em}{$\stackrel{\displaystyle <}{\sim}$}\;}

\def\ga{\gamma}

\def\si{\sigma}

\def\Ga{\Gamma}

\newcommand{\br}{{\rm BR}}

\newcommand{\hbb}{h \to b\bar{b}}

\newcommand{\htautau}{h \to \tau^+\tau^-}

\newcommand{\hWW}{h \to WW^*}

\newcommand{\hgaga}{h \to \ga\ga}

\def\3{\ss}

\begin{document}


\title{{\small{2005 International Linear Collider Workshop - Stanford,
U.S.A.}}\\ 
\vspace{12pt}
Exploring Complex Phases of the MSSM at Future Colliders} 


%




\author{S.~Heinemeyer}
\affiliation{CERN TH Division, Department of Physics,
CH-1211 Geneva 23, Switzerland}
\author{M.~Velasco}
\affiliation{Northwestern University, Evanston, Illinois 60201, USA}

\begin{abstract}
Once Supersymmetry is discovered, exploring the phases of
supersymmetric parameters will be one of the most important tasks of
future colliders. We analyze the possibilities of investigating the
phases of the cMSSM via their effects on the Higgs sector through
radiative corrections. Within two benchmark scenarios we compare the
capabilities of the LHC, the ILC and a future $\ga$C. 
\end{abstract}


\maketitle

\thispagestyle{fancy}







\section{INTRODUCTION} 

The Minimal Supersymmetric Standard Model (cMSSM) may possess several
complex phases. These phases can enter via loop corrections into the
Higgs boson sector~\cite{mhiggsCPXgen} and affect the Higgs boson
masses end 
couplings~\cite{mhiggsCPXEP,mhiggsCPXRG1,mhiggsCPXRG2,mhiggsCPXsn,mhiggsCPXFD1,mhiggsCPXFDproc}.
Most prominently the phase of the third
generation trilinear couplings, $\phi_{A_{t,b}}$, have an effect,
while the phases from the gaugino sector usually have a smaller
impact. At the two-loop level also the phase of the gluino,
$\phi_{\gl}$, can enter. Measuring these phases will be one of the
important tasks of future high-energy colliders. 

We discuss  the impact of complex phases within the
MSSM on various Higgs boson production and decay channels. Results are
compared for a the LHC, the ILC, and a 
$\ga\ga$~collider ($\ga$C). While the precision of the
branching ratio measurement at the LHC is not accurate enough, both
the ILC and the $\ga$C could in principle be sensitive to
the effects of complex phases (depending on the scenario). 
The precisions for the various Higgs boson decay channels at the three
colliders are summarized in tab.~\ref{tab:rates}. The Higgs boson mass
is set to ``typical'' values below the upper bound of \
$\mh \lsim 140 \gev$~\cite{mhiggslong,mhiggsAEC},
which is valid in the real as well as in the complex MSSM.

\begin{table}[htb!]
\renewcommand{\arraystretch}{1.3}
\begin{center}
\caption{Expected experimental precision of the rate measurement of
$h \to X$ at the LHC, the ILC operating at $\sqrt{s} = 500, 1000 \gev$, 
and the $\ga$C (based on the CLICHE design~\cite{cliche}).
}
\begin{tabular}{llcccccc}
\hline 
Study &  $\mh$ & $b \bar b$ & $W W^*$ & $\tau^+ \tau^-$ & $c \bar c$
                 & $gg$ & $\ga\ga$ \\
 \hline \hline
LHC ~\cite{zeppi}                                   & 120 GeV &
                 $\sim 20\%$  & $\sim 10\%$ & $\sim15\%$ & --- & ---   & --- \\
\hline 
ILC ($\sqrt{s} = 500 \gev$) ~\cite{tesla_ee,talkbrient} & 120 GeV & 
                  1.5\% & 3\% & 4.5\% & 6\% & 4\%   & 19\%  \\ \hline
ILC ($\sqrt{s} = 1000 \gev$) ~\cite{tesla_ee,barklow}  & 120 GeV & 
                  1.5\% & 2\% & ---   & --- & 2.3\% & 5.4\% \\ \hline
$\ga$C ~\cite{cliche,LeptonPhoton}                  & 115 GeV  
                 & 2\% & 5\% &  ---  & --- & ---   & 22\%  \\ \hline
\hline
\end{tabular}
\label{tab:rates}
\end{center}
\end{table}


\section{COMPARISON OF DIFFERENT COLLIDERS}

We compare the sensitivity of a future $\ga$C
with that of the LHC and the ILC. The comparison is based on two
different physics scenarios:\\

\newpage
\noindent
\underline{The CPX scenario:}\\
This scenario has been designed to give maximum effects of
CP-violating phases~\cite{cpx}. The parameters are
\BEA
&& \Msusy = 500 \gev, |\At| = 1000 \gev, \At = \Ab = A_\tau \non \\
\label{cpxpar}
&& M_2 = 500 \gev, |\mgl| = 1000 \gev, \mu = 2000 \gev\\
&& \phi = \phi_{A_{t,b,\tau}} = \phi_{\mgl} \non
\EEA
$\Msusy$ denotes a common soft SUSY-breaking mass in the sfermion mass
matrices. $A_f$ is the trilinear Higgs-Sfermion coupling with the
phase $\phi_f$. $M_2$ is a gaugino mass parameter, $\mgl$ denotes the
gluino mass, and $\mu$ is the Higgs mixing parameter.

\noindent
\underline{The BGX scenario:}\\
This scenario is motivated by baryogenesis. It has been shown in
\cite{bgx} that in this scenario (depending on the Higgs sector
parameters) baryogenesis in the early universe could be possible. It
is thus a physics motivated scenario, not emphazising possible effects
of complex phases. The parameters are
\BEA
&& \MstL = 1.5 \tev, \MstR = 0, M_{\tilde Q_{1,2}} = 1.2 \tev,
   M_{\tilde L_{1,2}} = 1.0 \tev \non \\
&& |\Xt| = 0.7 \tev, \At = \Ab = A_\tau \non \\
\label{bgxpar}
&& M_2 = 220 \gev, \mgl = 1 \tev, \mu = 200 \gev\\
&& \phi = \phi_{A_{t,b,\tau}} = \phi_{\mgl} \non
\EEA
Here $M_{\tilde t_{L,R}}$ are the soft SUSY-breaking parameters in the
scalar top mass matrix. $M_{\tilde Q_{1,2}}$ are the corresponding
parameters for the squarks of the first two generations, while 
$M_{\tilde L_{1,2}}$ refer to the sleptons of the first two
generations. $m_t \Xt$ is the off-diagonal entry in the scalar top
mass matrix with $\Xt = \At - \mu/\tb$.

The results presented here have been obtained with the code 
{\em FeynHiggs2.2}~\cite{feynhiggs,mhiggslong,mhiggsAEC,mhiggsCPXFD1}. 
It should be noted that
the higher-order uncertainties in these evaluations are somewhat less
under control as compared to the real case, see e.g.\
\citere{habilSH}. The same applies to the parametric uncertainties due
to the experimental errors of the input
parameters~\cite{habilSH,tbexcl,mhiggsWN,PomssmRep}. 
Results for branching ratios 
obtained with an alternative code, {\em CPsuperH}~\cite{cpsh}, can
differ quantitatively to some extent from the results shown here. 
A main difference between the two codes are the more complete inclusion
of real two-loop corrections in {\em FeynHiggs2.2}, resulting in
somewhat higher values for the lightest Higgs boson mass. While the
complex phase dependence at 
the one-loop level is included completely in {\em FeynHiggs2.2}, at
the two-loop level it is more complete in {\em CPsuperH}, which makes
it difficult to disentangle the source of possible deviations. A more
complete discussion can be found in \cite{FHCPsHcomp}.


\subsection{The CPX scenario}

\begin{figure}[ht] 
     \begin{center}
     \epsfig{file=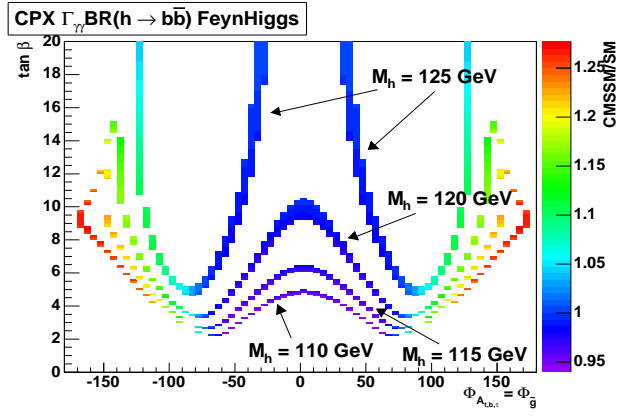,width=8.40cm,height=5.50cm}
     \end{center}
\vspace{-0.5em}
     \caption{
The deviations of $\Ga_{\ga\ga} \times \br(\hbb)$ within the CPX scenario
from the SM value is shown in the $\phi_{A_{t,b}}$--$\tb$ plane.
The corresponding precision obtainable at a $\ga$C is $\sim 2\%$.
     } 
     \label{fig:gg_cpx}
     \end{figure}

We start our analysis by the investigation of the CPX scenario, see
eq.~\ref{cpxpar}. 
We first show the results for the $\ga$C in fig.~\ref{fig:gg_cpx} for the
decay channel $\hbb$, which has the best sensitivity at this
collider. The variation of $\Ga_{\ga\ga} \times \br(\hbb)$ is shown in
the $\phi_{A_{t,b}}$--$\tb$ plane. The strips correspond to constant
values of the lightest Higgs mass, while the color code shows the
deviation from the corresponding SM value. It should be kept in mind
that the Higgs boson mass will be measured to very high accuracy so
that one will be confined to one of the strips. 
We are neglecting the parametric errors from the imperfect knowledge
of the input parameters. In reality these parametric errors would
widen the strips. The future intrinsic error of 
$\sim 0.5 \gev$~\cite{PomssmRep}, 
however, is included in the width of the strips.
One can see that this channel can be strongly enhanced as compared
to the SM. The variation along each strip is much larger than
the anticipated precision of $\sim 2\%$ for this channel. This would
allow to constrain the values of the complex phases. The picture
becomes of course more complicated if the complex phases are varied
independently. Various channels will have to be combined to disentangle
the different effects.

%
     \begin{figure}[htb!]
     \begin{center}
     \begin{tabular}{cc}
     \mbox{\epsfig{file=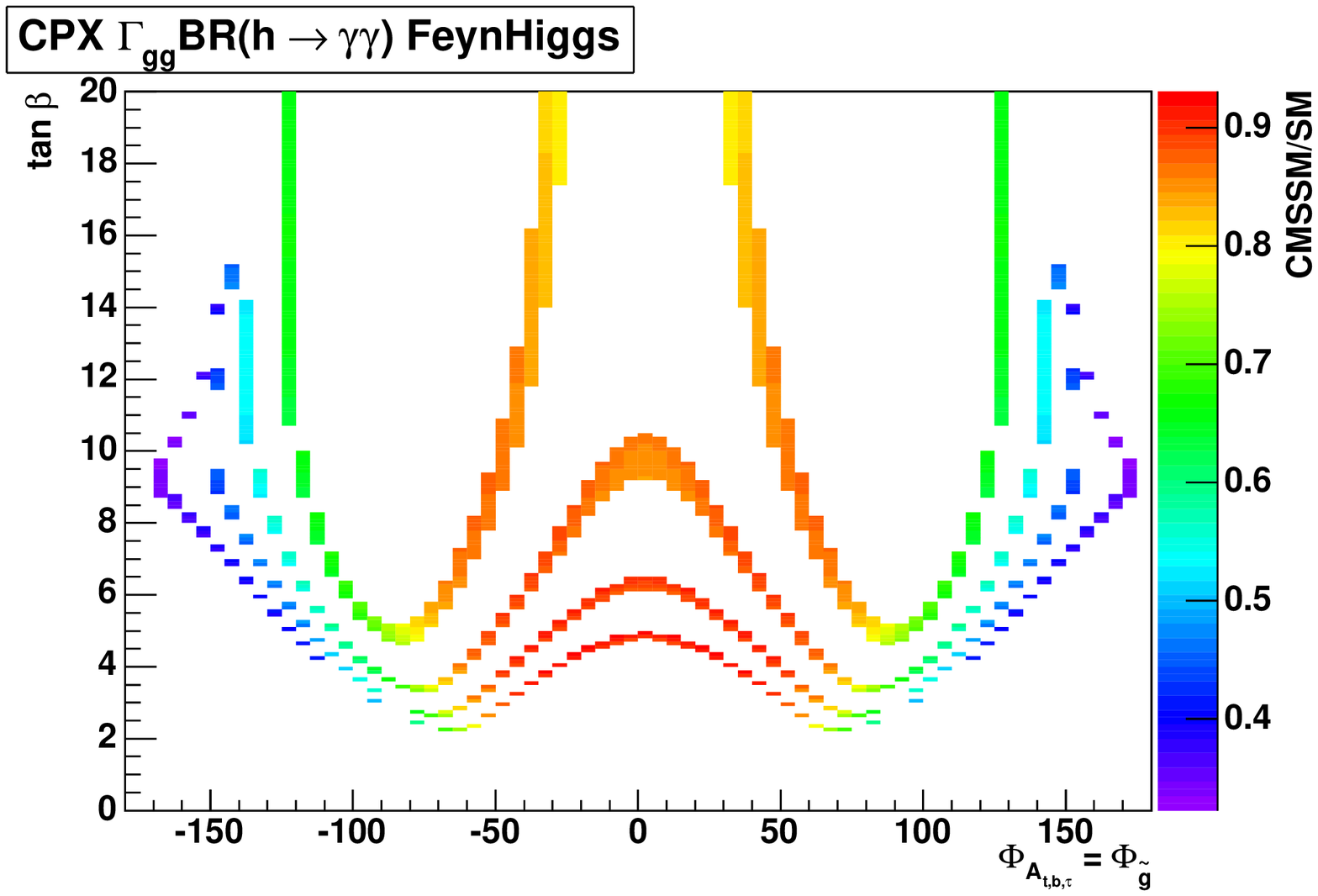,width=8.40cm,height=5.50cm}}&
     \mbox{\epsfig{file=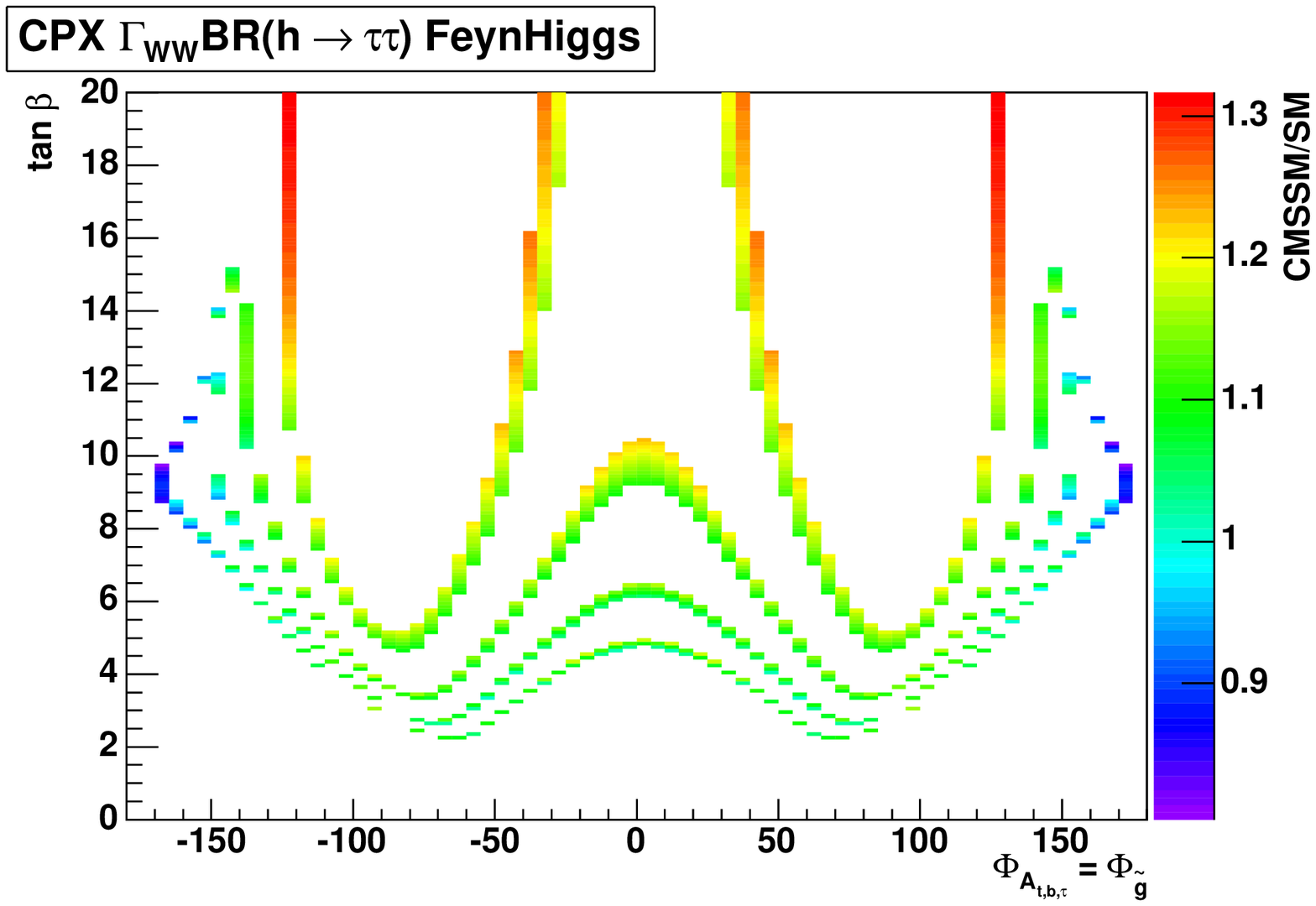 ,width=8.40cm,height=5.50cm}}
     \end{tabular}
     \end{center}
\vspace{-0.5em}
     \caption{
The deviations of $\Ga_{gg} \times \br(\hgaga)$ (left) and of 
$\Ga{WW} \times \br(\htautau)$ (right) within the CPX scenario
from the SM value is shown in the $\phi_{A_{t,b}}$--$\tb$ plane.
The corresponding experimental precision can be found in
tab.~\ref{tab:rates}. 
}
     \label{fig:lhc_cpx}
     \end{figure}
%
     \begin{figure}[hb!]
     \begin{center}
     \begin{tabular}{cc}
     \mbox{\epsfig{file=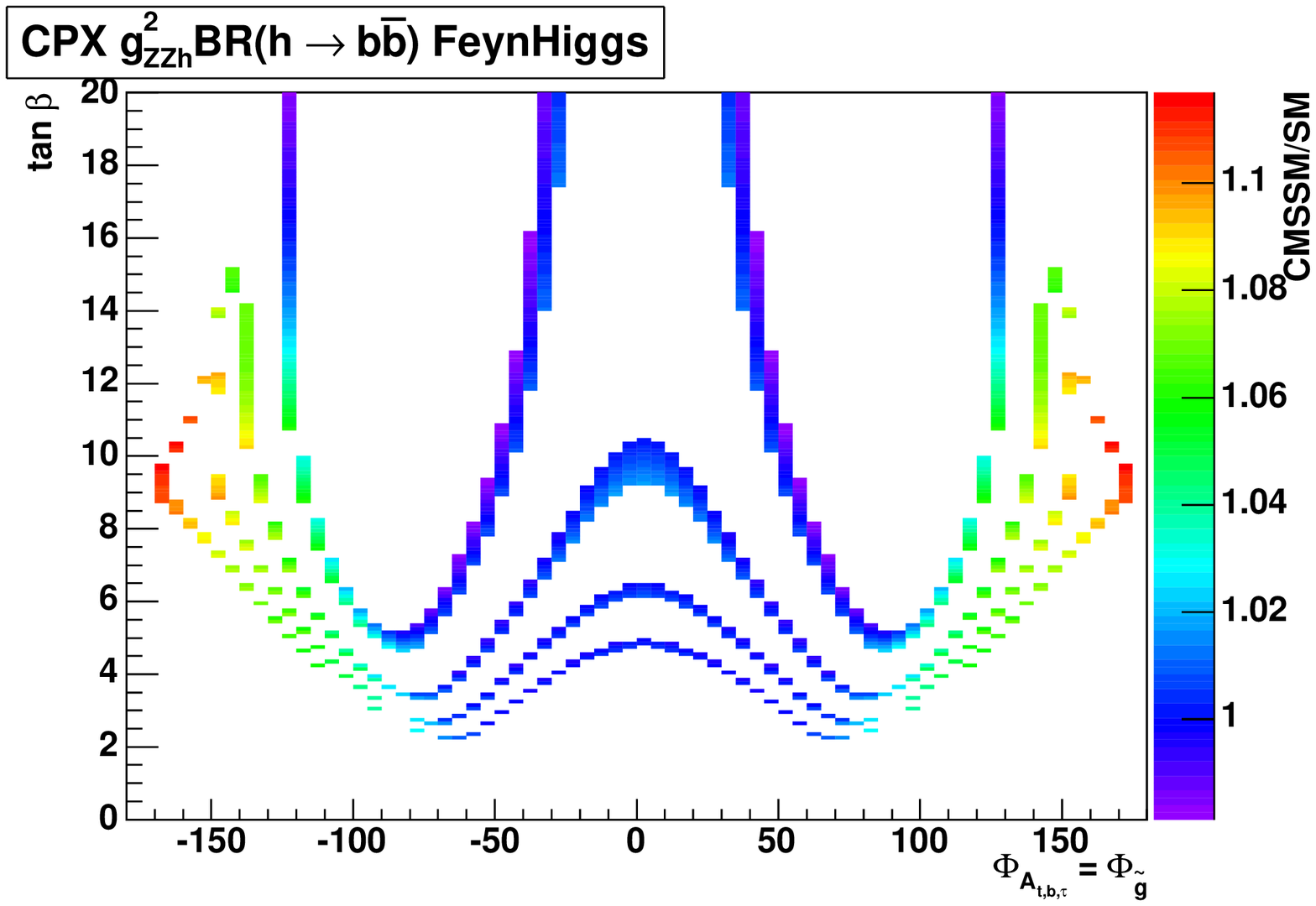,width=8.40cm,height=5.50cm}}&
     \mbox{\epsfig{file=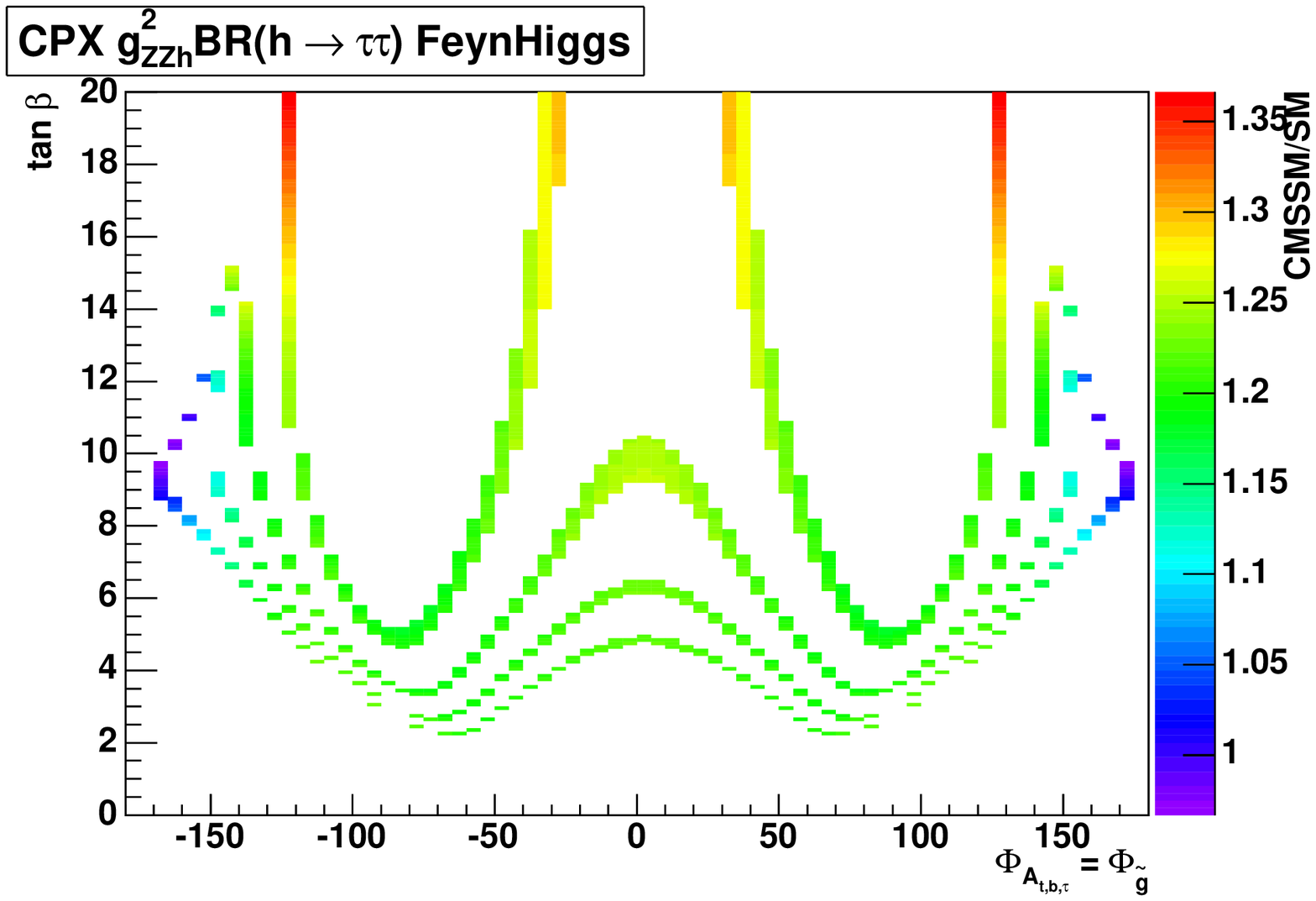 ,width=8.40cm,height=5.50cm}}
     \end{tabular}
     \end{center}
\vspace{-0.5em}
     \caption{
The deviations of $g_{ZZh}^2 \times \br(\hbb)$ (left) and of 
$g_{ZZh}^2 \times \br(\htautau)$ (right) within the CPX scenario
from the SM value is shown in the $\phi_{A_{t,b}}$--$\tb$ plane.
The corresponding experimental precision can be found in
tab.~\ref{tab:rates}. 
}
     \label{fig:ilc_cpx}
     \end{figure}

Results for the LHC are shown in fig.~\ref{fig:lhc_cpx}. The left
plots gives the results for the channel $gg \to h \to \ga\ga$, while
the right plots depicts $WW \to h \to \tau^+\tau^-$.
The latter channel (like $h \to b \bar b$) is usually somewhat
enhanced in the MSSM, the $\br(\hWW)$ (not shown) and
$\br(\hgaga)$ (see the left plot of \ref{fig:lhc_cpx})
are normally suppressed in this scenario. The precision of
the LHC will not be good enough to obtain information about complex
phases in this way. 

Finally in fig.~\ref{fig:ilc_cpx} shows the ILC results in the CPX
scenario. The left plot shows the BR($h \to b \bar b$), while the
right plot depicts BR($h \to \tau^+\tau^-$). Both channels are
enhances as compared to the SM in this scenario.
The high precision of the ILC (see \ref{tab:rates})
shows that this collider has 
a good potential to disentangle the complex phases.

Since in the examples shown here for the $\ga$C and the ILC 
the largest deviations occur for
different regions of the parameter space, the results from both
colliders could be combined in order to extract the maximum
information on $\phi_{A_{t,b}}$.


\subsection{The BGX scenario}

Now we turn to the investigation of the baryogenesis motivated
BGX scenario, see eq.~\ref{bgxpar}. The effects in this scenario are
expected to be smaller than in the CPX scenario that had been designed
to give maximum effects of the complex phases. 

In fig.~\ref{fig:gg_bgx} we show the $h \to b \bar b$ channel at the
$\ga$C. A substantial suppression with respect to the SM can be
observed. However, the variation of 
$\Ga_{\ga\ga} \times {\rm BR}(h \to b \bar b)$ for fixed Higgs boson
mass (which will be known with high precision) with the complex phase
$\phi_{A_{t,b}}$ is very small. Thus a precise measurement of this
channel at the $\ga$C will not reveal any information about the
complex phases entering the MSSM Higgs sector. 

The two LHC channels in the BGX scenario are shown in
fig.~\ref{fig:lhc_bgx}, while the two ILC channels are given in
fig.~\ref{fig:ilc_bgx}. As for the CPX scenario no phase measurement
can be expected from he LHC measurements. The situation at the ILC in
the BGX scenario is similar to the $\ga$C. A deviation from the SM
value can be measured, but the variation of 
$g_{ZZh}^2 \times {\rm BR}(h \to b \bar b, \tau^+\tau^-)$ 
is too small to reveal any information on $\phi_{A_{t,b}}$.


\section{CONCLUSIONS}

We have compared the LHC, the ILC and the $\ga$C in view of their
power to determine the complex phases of the cMSSM. We have focused on
the Higgs sector, where the complex phases enter via radiative
corrections. Especially we have investigated the most promising
combinations of Higgs production and decay ($\si \times \br$) for each
collider.

The analysis has been performed in two scenarios:
The CPX scenario designed to maximize the effect of
complex phases in the MSSM Higgs sector.
The other scenario (BGX) is based on a part of the cMSSM that is motivated
by baryogenesis.

The CPX scenario may offer good prospects for
the $\ga$C and the ILC to determine $\phi_{A_{t,b}}$ via Higgs
branching ratio measurements. On the other hand, the BGX
scenario will only show a deviation from the
SM. The variation of the analyzed channels is too small to give
information on the complex phases.

\begin{figure}[ht] 
     \begin{center}
     \epsfig{file=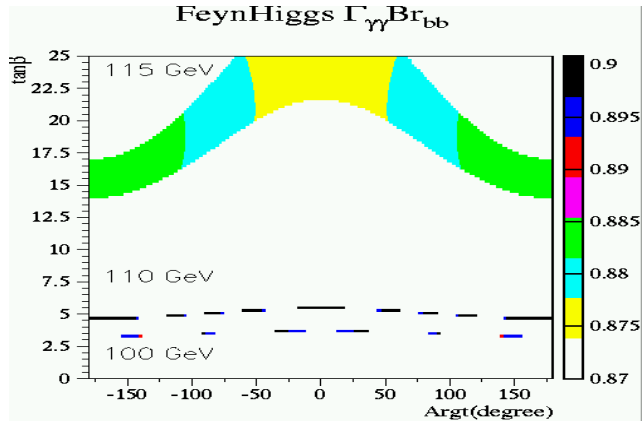,width=8.40cm,height=5.5cm}
     \end{center}
\vspace{-0.5em}
     \caption{
The deviations of $\Ga_{\ga\ga} \times \br(\hbb)$ within the BGX scenario
from the SM value is shown in the $\phi_{A_{t,b}}$--$\tb$ plane.
The corresponding precision obtainable at a $\ga$C is $\sim 2\%$.
     } 
     \label{fig:gg_bgx}
     \end{figure}

\begin{figure}[htb!]
     \begin{center}
     \begin{tabular}{cc}
     \mbox{\epsfig{file=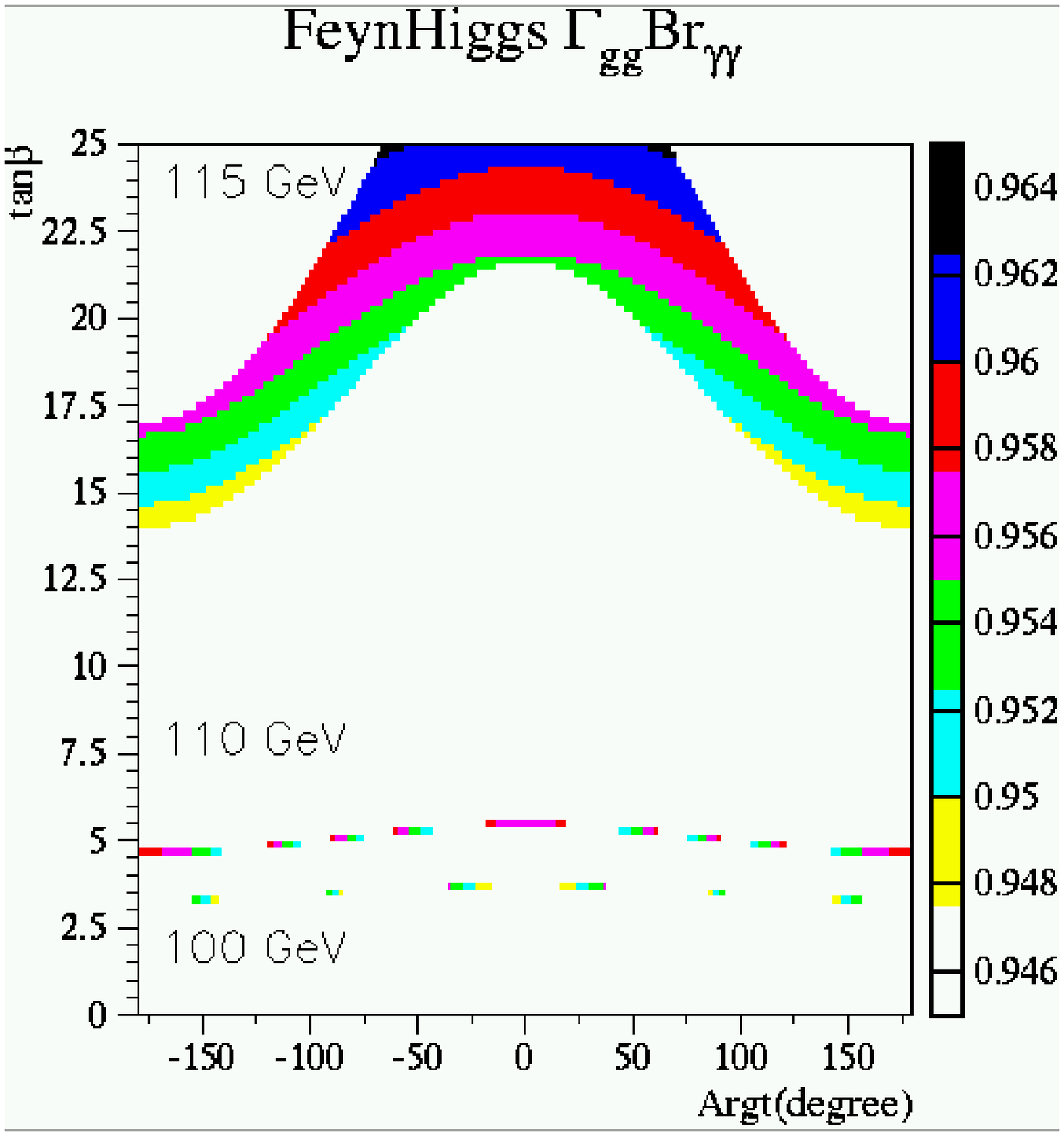,width=8.40cm,height=5.5cm}}&
     \mbox{\epsfig{file=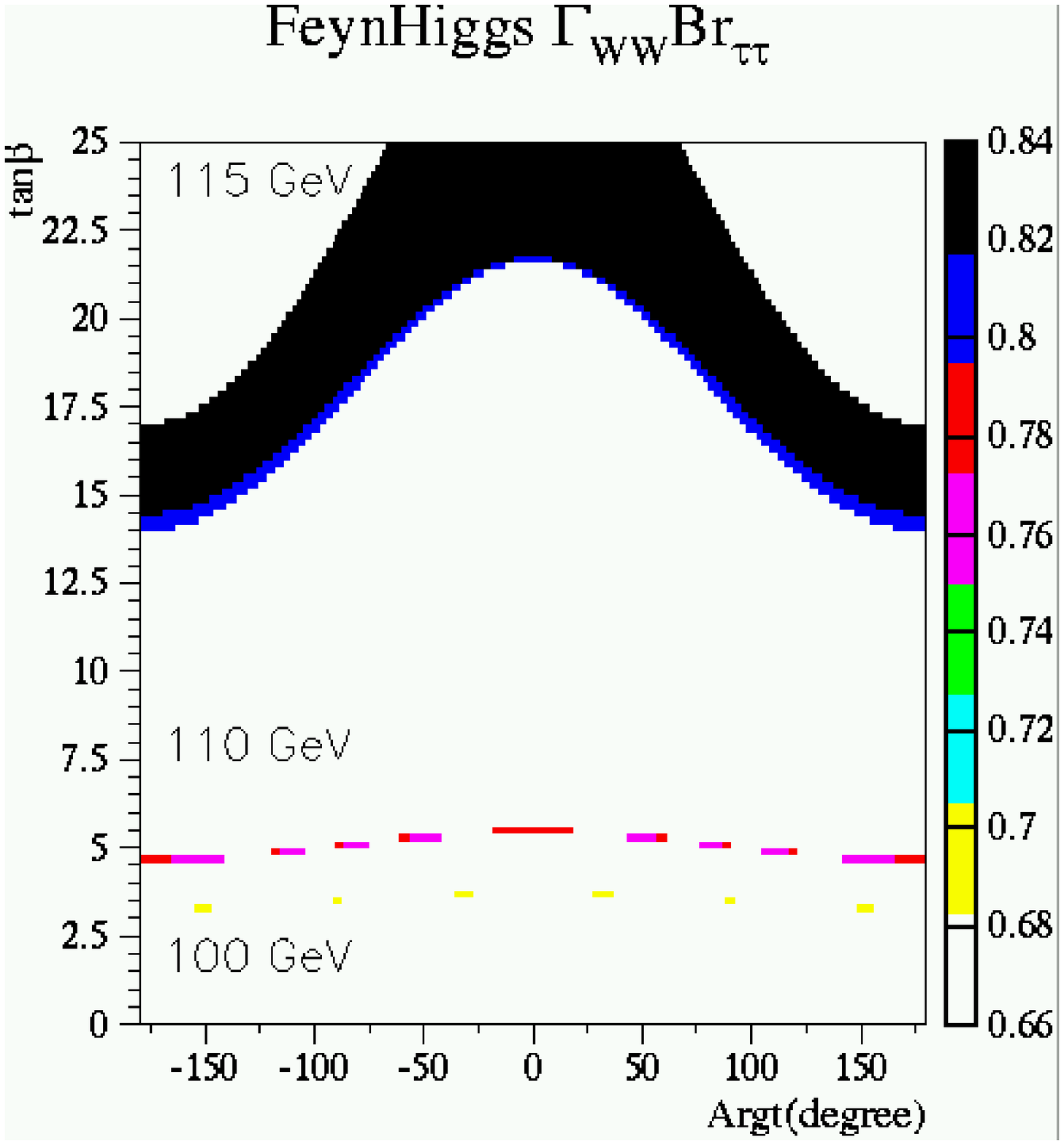 ,width=8.40cm,height=5.5cm}}
     \end{tabular}
     \end{center}
\vspace{-0.5em}
     \caption{
The deviations of $\Ga_{gg} \times \br(\hgaga)$ (left) and of 
$\Ga{WW} \times \br(\htautau)$ (right) within the BGX scenario
from the SM value is shown in the $\phi_{A_{t,b}}$--$\tb$ plane.
The corresponding experimental precision can be found in
tab.~\ref{tab:rates}. 
}
     \label{fig:lhc_bgx}
     \end{figure}

\begin{figure}[htb!]
     \begin{center}
     \begin{tabular}{cc}
     \mbox{\epsfig{file=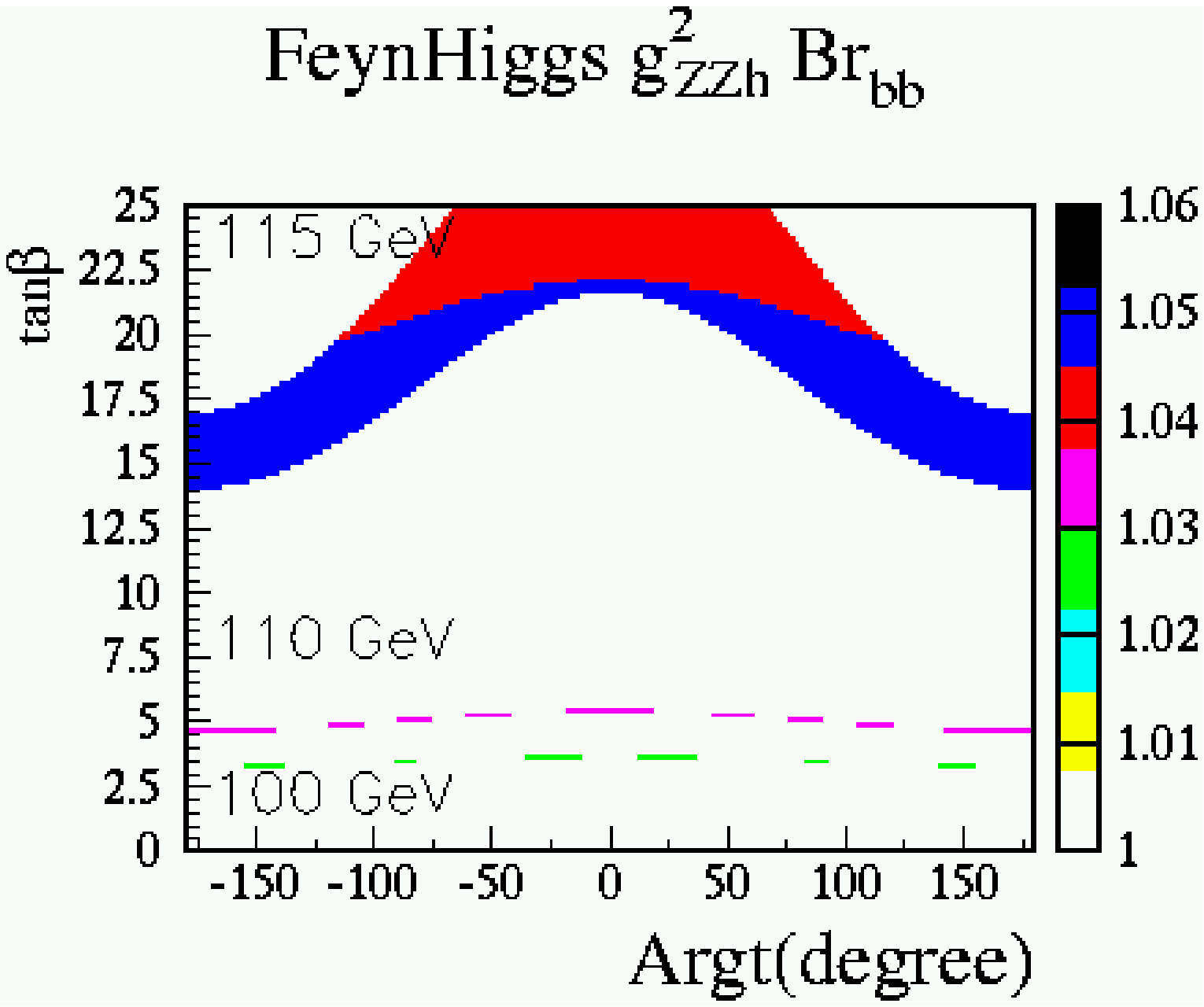,width=8.40cm,height=5.5cm}}&
     \mbox{\epsfig{file=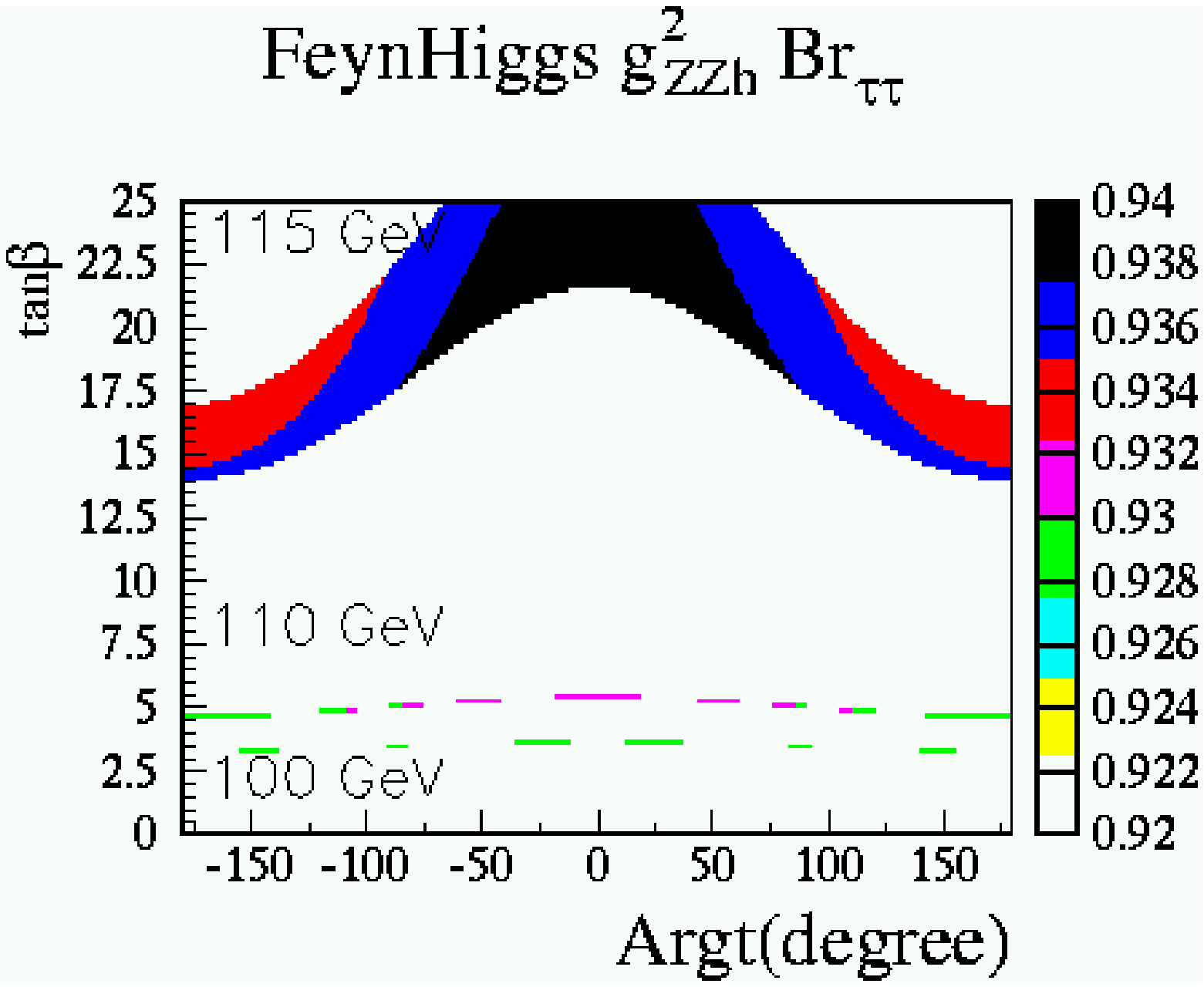 ,width=8.40cm,height=5.5cm}}
     \end{tabular}
     \end{center}
\vspace{-0.5em}
     \caption{
The deviations of $g_{ZZh}^2 \times \br(\hbb)$ (left) and of 
$g_{ZZh}^2 \times \br(\htautau)$ (right) within the BGX scenario
from the SM value is shown in the $\phi_{A_{t,b}}$--$\tb$ plane.
The corresponding experimental precision can be found in
tab.~\ref{tab:rates}. 
}
     \label{fig:ilc_bgx}
     \end{figure}

It should be kept in mind that we have neglected the future parametric
errors on the SUSY parameters (see e.g.\ \citere{lhcilc} and
references therein). These uncertainties will further widen
the bands shown in figs.~\ref{fig:gg_cpx}--\ref{fig:ilc_bgx}.


\subsection*{Acknowledgments}

S.H. thanks C.~Wagner for helpful discussions concerning the BGX scenario.

\newpage

\psfrag{Argt(degree)}[][]{$\phi_{A_{t,b}}$}
%
%




\end{document}